\documentclass[letter]{aa}

\usepackage[normalem]{ulem}
\usepackage{xcolor}
\usepackage{graphicx,txfonts,textcomp}
\usepackage{natbib,twoopt}
\usepackage{placeins}
\usepackage{hyperref}
\usepackage{xcolor}
\usepackage{lineno}  
\modulolinenumbers[1]  

\authorrunning{Chen et al.}

\begin{document}

	\title{A Short-timescale Negative Optical Continuum Lag in SDSS J083717.88+191647}
	
	\author{Mu-Lin Chen\inst{1} \and Pei-Zhen Cheng\inst{1} \and Xing-Qian Chen\inst{1} \and Ying Gu\inst{1} \and Qi Zheng\inst{1} \and Gui-Lin Liao\inst{1} \and Xue-Guang Zhang\inst{1}\fnmsep\thanks{E-mail: {xgzhang@gxu.edu.cn}} 
	}
	
	\institute{Guangxi Key Laboratory for Relativistic Astrophysics, School of Physical Science and Technology, GuangXi University, No. 100, Daxue Road, Nanning 530004, P. R. China
	}
	
	\date{Received xxx; accepted xxx}

	\abstract
	{Continuum reverberation mapping (RM) is a powerful technique for constraining the accretion disk structure in active galactic nuclei (AGNs). In typical cases, the shorter-wavelength emission is used as the reference, and a positive time lag is observed since the inner, hotter regions of the accretion disk respond earlier than the cooler outer regions at longer wavelengths. However, we detect a short-timescale negative inter-band lag in SDSS~J083717.88+191647 using RM techniques, where the \textit{g}-band lags behind the \textit{r}-band emission. The light curves from the Zwicky Transient Facility reveal two distinct phases, a stabilizing and a declining phase, in which the time lags show opposite signs. Using \texttt{JAVELIN} with the $g$-band as the reference, we obtain time lags of $3.68^{+1.94}_{-2.78}$~days during the stabilizing phase and $-1.60^{+0.69}_{-0.54}$~days during the declining phase. Although negative continuum lags have been reported in a few previous studies, the present case is distinguished by its clear phase dependence and the accompanying color evolution. We attribute the observed lag reversal to a moving dust-cloud obscuration scenario, in which the cloud crossing the line of sight preferentially obscures emission from the outer longer-wavelength regions of the disk, causing the $r$-band to decline earlier than the $g$-band and thus producing the observed negative inter-band lag. Our results indicate that AGN variability may be more complex than previously thought. Future high-cadence, multi-band observations will be essential to test this dust-obscuration model and to further explore the interplay between the accretion disk emission and dust in AGNs.
	}
	
	
	\maketitle
	%
	
	\section{Introduction}
	
	Continuum reverberation mapping (RM), first proposed by \citet{blandford1982}, is a powerful approach to constraining the disk structure of active galactic nuclei (AGNs). By measuring inter-band time delays in continuum variations, this technique allows one to estimate the characteristic size of the accretion disk \citep{1993PASP..105..247P}. Within the framework of the standard Shakura-Sunyaev (SS) disk model \citep{Shakura1973}, shorter-wavelength emission originating from the hotter, inner regions is expected to lead the longer-wavelength emission produced at larger radii, resulting in observed positive time lags \citep{2007A&A...467.1057T,2012JPhCS.372a2058G} relative to the shorter-wavelength reference.
	
	Over the past decade, continuum RM has provided robust measurements of positive inter-band time lags across more than 200 AGNs, with delays typically spanning a few hours to $\sim$10 days \citep{2018ApJ...857...53C,2021iSci...24j2557C,2023ApJ...953..137M, 2025MNRAS.543.2093F}. 
	\citet{2023A&A...672A.132F} find positive delays for MCG 08-11-011, with lags of 1.0 $\pm$ 0.5, 2.0 $\pm$ 0.6, 4.5 $\pm$ 0.6, and 7.1 $\pm$ 1.1 days across bands from 4250 to 8025 Å, yielding results 3--7 times larger than SS model predicted. Similar continuum RM studies \citep[e.g.,][]{2017FrASS...4...55F,2018ApJ...854..107F, 2022MNRAS.511.3005J, 2024ApJ...966..149J, 2025A&A...695A.143W} have also found disk sizes larger than SS-model predictions, possibly due to broad-line region (BLR) diffuse continuum contamination, scattering, or geometric effects. Beyond disk size inferences, continuum RM leveraging these positive time lags has broadened to other AGN applications, including black hole (BH) mass estimation \citep{2020ApJS..246...16Y, 2023ApJ...948L..23W}, refined scaling relations \citep{2022MNRAS.509.2637N, 2025ApJ...985...30M}, disk--broad-line region interactions \citep{2022ApJ...925...29C, 2024ApJ...968L..16P, 2024ApJ...974..288Z,2025A&A...702A..92J} and so on. 
	
	While previous studies have predominantly focused on positive inter-band lags in AGNs, some intriguing exceptions have emerged. \citet{Guo2022} have found that negative $g-r$ lags can appear in a small subset of Zwicky Transient Facility (ZTF) AGNs. Negative-lag behavior has also been reported in other AGN lag studies \citep{2025ApJ...993..245L}. In Fairall~9, \citet{yao2023} have interpreted the long-timescale negative lag in terms of changes in the accretion-disk vertical structure, and \citet{2023ApJ...956...81S} have shown through simulations that similar behavior can also arise from structural changes in the accretion flow. These results indicate that negative continuum lags are not unique, although their physical origin may be different.
	
	\begin{figure*}[ht!]
		\centering
		\includegraphics[width=0.62\textwidth]{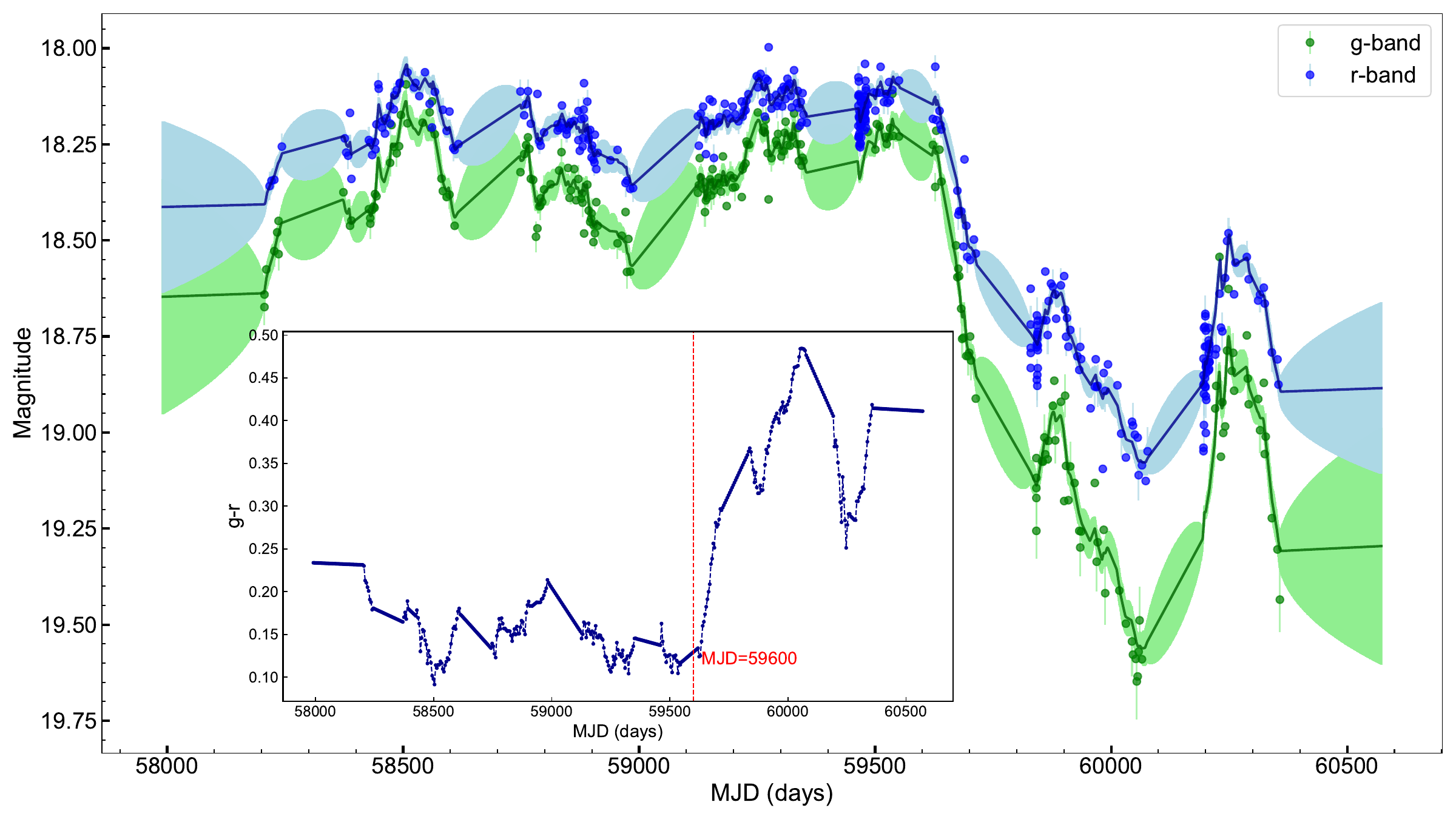}
		\includegraphics[width=0.365\textwidth]{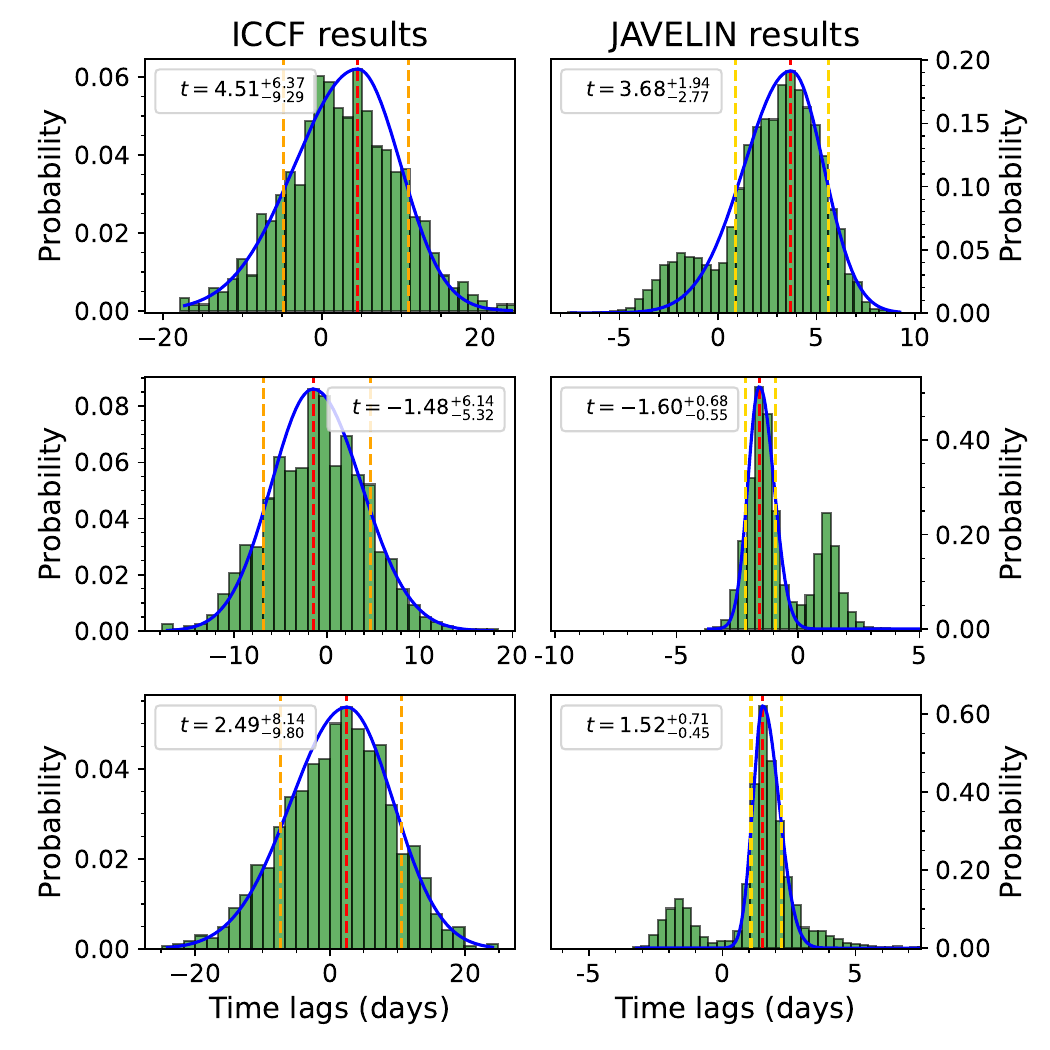}
		\caption{Left panel: Light curves of SDSS J0837 in $g$- (green dots) and $r$-bands (blue dots) overlaid with best-fit (green and blue shadows) \texttt{JAVELIN} models. The inset shows the $g-r$ color evolution, with a vertical line marking MJD 59600. 
			Right panels: Time lag histograms for the stabilizing (top), declining (middle), and entire (bottom) phases. Green bars show the posterior distributions with Gaussian-like fits (blue curves), while red and orange dashed lines indicate the peak positions and $1\sigma$ uncertainties, respectively.
		}
		
		\label{fig:light-curve}
	\end{figure*}

	In this work, we report a short-timescale negative lag between the $g$- and $r$-bands in SDSS J083717.88+191647 (hereafter SDSS~J0837) using ZTF data. Specifically, during the stabilizing phase, the $g$-band variations lead the $r$-band showing a conventional positive lag, whereas during the declining phase, a negative lag is observed, with the $g$-band variations lagging behind the $r$-band using the $g$-band as a reference. SDSS~J0837 therefore provides a clear example of a phase-dependent lag reversal. To interpret this observation, a moving dust-cloud obscuration scenario is proposed. The remainder of this paper is organized as follows: Section \ref{data and results} describes the data and results; Section \ref{con} summarizes the conclusion. Throughout the manuscript, we have adopted the cosmological parameters
	of $H_{0}$=70 km s$^{-1}$ Mpc$^{-1}$, $\Omega_{m}$=0.3, and $\Omega_{\Lambda}$=0.7.
	
	\section{Data and Results} \label{data and results}
	
	The photometric data of SDSS J0837 ($z=0.293$) are obtained from the ZTF survey \footnote{\url{https://irsa.ipac.caltech.edu/cgi-bin/Gator/nph-dd}} \citep{2019PASP..131a8003M,2019PASP..131a8002B}, comprising 232 and 363 epochs in the $g$- and $r$-bands (Figure~\ref{fig:light-curve}, left panel), respectively, over the period MJD 58206--60357. 
	Only measurements with ${\tt catflags}=0$ and ${\tt programid}=1$ are retained to exclude problematic points and avoid non-uniform sampling.
	The $i$-band data are excluded due to sparse sampling and potential H$\alpha$ contamination \citep{2023ApJ...953..137M, 2025ApJ...985...30M}. To investigate the source evolution, we examine the $g-r$ color evolution (Figure~\ref{fig:light-curve}, left panel inset). The color index remains nearly constant before MJD 59600, indicating a relatively stable phase. Thereafter, pronounced reddening and strong color evolution emerge, most likely associated with dust obscuration, underscoring the unusual nature of the source. 
	Dust contamination at longer wavelengths has been reported in Fairall 9 \citep{Mandal:2025fnm}, where positive lags are attributed to disk and dust reprocessing, supporting a different origin for the observed negative lag here.
	We therefore separate the variability into two distinct phases: a stabilizing phase before MJD 59600 and a declining phase thereafter. This clear phase division, together with the contemporaneous $g-r$ color evolution, makes SDSS~J0837 a useful case for investigating the origin of the lag inversion.
	
	We apply \texttt{JAVELIN} \citep{2011ApJ...735...80Z,2013ApJ...765..106Z, 2016ApJ...819..122Z} and interpolated cross-correlation function (ICCF) \citep{1986ApJ...305..175G, 1987ApJS...65....1G,1998PASP..110..660P} to measure inter-band time lags in different phases of the ZTF light curves. The \texttt{JAVELIN} best-fit results are illustrated by the green and blue shadows in Figure~\ref{fig:light-curve}, left panel. During the stabilizing phase, \texttt{JAVELIN} detects a positive lag where the $g$-band leads the $r$-band by $3.68^{+1.94}_{-2.78}$ days. However, this behavior reverses during the declining phase, yielding a lag of $-1.60^{+0.69}_{-0.54}$ days, which implies that the $r$-band variations precede those in the $g$-band. Analysis of the entire light curve yields an intermediate lag of $1.52^{+0.71}_{-0.45}$ days. The ICCF results are qualitatively consistent with this trend, showing lags of $4.51^{+6.37}_{-9.28}$ days, $-1.48^{+6.13}_{-5.32}$ days, and $2.49^{+8.14}_{-9.80}$ days for the stabilizing, declining, and full phases, respectively. Although the ICCF results show larger uncertainties, likely owing to its sensitivity to irregular sampling, both methods consistently support a transition from a positive to a negative time lag. Given that the cadence of the ZTF light curves is comparable to the measured lag and includes significant gaps, we assess the impact of the sampling pattern through simulations, finding that sparse sampling and large gaps in the ZTF data do not produce spurious negative lags (see Appendix~\ref{apenb}).

	Motivated by the lag reversal and the $g-r$ color evolution, we propose a moving dust-cloud obscuration scenario. In this picture, a dust cloud crosses the line of sight and progressively obscures the accretion disk. Because the disk emission is radially stratified, the cloud first occults the outer $r$-band emitting regions before reaching the inner $g$-band regions. During the stabilizing phase, when the disk is unobscured, the light curves display the expected positive time lag. In the subsequent ``outside-in'' occultation phase, however, a geometric time offset is introduced, such that the $r$-band flux declines before the $g$-band flux, naturally giving rise to the observed negative lag.
	
	
	To model this process, we simulate the intrinsic disk variability using a CAR(1) stochastic process \citep[e.g.,][]{2009ApJ...698..895K},
	\begin{equation}
		dm(t) = -\frac{1}{\tau} m(t)\,dt + \sigma \sqrt{dt}\,\epsilon(t) + m_0,
	\end{equation}
	where $\tau$ is the damping timescale, $\sigma$ is the variability amplitude, $m_0$ is the mean magnitude, and $\epsilon(t)$ is a Gaussian white-noise process. For the intrinsic $g$-band light curve, we adopt representative parameters of $\tau = 1000$~days, $\sigma = 0.01$~mag~day$^{-1/2}$, and $m_0 = 18.34$~mag, where the mean magnitude is taken from the average value during the stabilizing phase. We do not adopt the \texttt{JAVELIN} outputs inferred from the observed light curves, as they may be contaminated by the obscuration event. The intrinsic $r$-band light curve is then constructed from the intrinsic $g$-band variability through a Gaussian convolution and a magnitude renormalization. Specifically, we convolve the intrinsic $g$-band light curve with a Gaussian transfer function and impose an intrinsic time delay of $t_{\rm lag} \approx 3.68$~days, consistent with the \texttt{JAVELIN} measurement in the stabilizing phase. Since the \texttt{JAVELIN} analysis gives a transfer-function width close to zero (Figure~\ref{fig:width}), we adopt a narrow kernel with a width of 0.1~days as a numerical approximation to a $\delta$-function response. After the convolution, we subtract the mean magnitude of the $g$-band and add that of the $r$-band, adopting $\langle m_r \rangle = 18.18$~mag, also derived from the stabilizing phase.

	\begin{figure}
		\centering
		\includegraphics[width=\columnwidth]{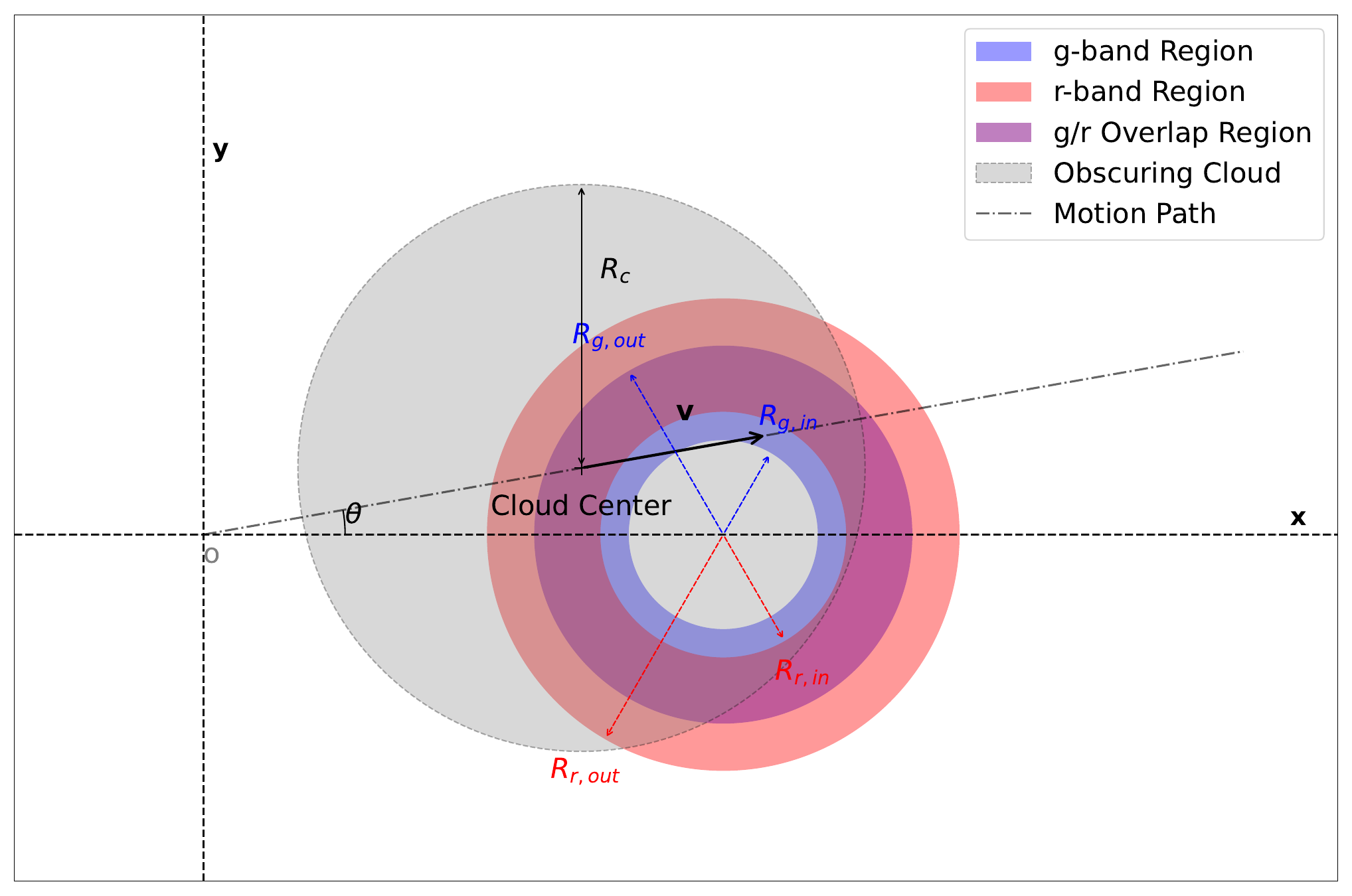} 
		\caption{Schematic of the obscuration model. The accretion disk comprises concentric $g$- (blue, $R_g$) and $r$-band (red, $R_r$) annuli, with their overlap (purple) representing diffuse continuum blending. The obscuring cloud ($R_c$, gray dashed circle) moves with velocity $v$ along a linear path inclined at angle $\theta$ relative to the disk plane (black dot-dashed line).
		}
		\label{fig:schematic}
	\end{figure}

	To simulate the obscuration event, we construct a geometric grid model of the emitting regions, as illustrated schematically in Figure~\ref{fig:schematic}. The model includes concentric emitting regions for the $g$- and $r$-bands, with a moving obscuring cloud. The $g$-band emission extends from $R_{g,in}=1.0$ to $R_{g,out}=2.0$, while the $r$-band spans $R_{r,in}=1.3$ to $R_{r,out}=2.5$. These emission regions are assumed to be uniform and partially overlapping, which provides a more realistic representation of the emitting structure \citep[e.g.,][]{Fausnaugh_2016, 2019ApJ...870..123E, 2018MNRAS.481..533L, 2021iSci...24j2557C}. To connect the model to physical scales, we adopt a spatial conversion in which 1 unit corresponds to 2.0~light-days. This choice is motivated by the BH mass of SDSS J0837 ($10^{9.02} M_\odot$; \citealt{2011ApJS..194...45S}), for which $1\,R_g \approx 1.37$~light-hours, and ensures that the optical emission regions lie within $\sim60$--$100\,R_g$, consistent with expectations for such a system. Under this scaling, the adopted radial ranges correspond to 2.0--4.0~light-days for the $g$-band and 2.6--5.0~light-days for the $r$-band.

	As shown in Figure~\ref{fig:schematic}, the obscuring cloud is modeled as a circular, semi-transparent structure with a radius of $R_c = 3.0$ units ($6.0$ light-days) and a center at $(x_c, y_c)$. To account for inhomogeneity, we adopt a Gaussian spatial distribution for the color excess $E(B-V)$, representing a centrally concentrated cloud with decreasing density toward the edges. The accretion disk is centered at $(R_c + R_{r,\rm out},\, 0)$, such that the cloud is initially externally tangent to the outer boundary of the disk emitting region. The cloud then moves across the disk over 1000 days along a straight trajectory inclined by $\theta = 10^{\circ}$ to the $x$-axis. The disk emission is discretized on a two-dimensional Cartesian grid with $i \times j = 400 \times 400$ cells, where $(i,j)$ denote the grid indices and $(x_i, y_j)$ the corresponding physical coordinates. At each time step $t$, the intrinsic magnitudes $m_{\rm int}(t)$ are converted to fluxes according to $F_{\rm int}(t) = F_0\,10^{-0.4\, m_{\rm int}(t)}$, where $F_0$ is the zero-point flux. Since $F_0$ only provides an overall normalization, we set $F_0 = 1$. The resulting flux is then uniformly distributed over the corresponding annulus, such that $F_{\rm pixel}(i,j,t) = F_{\rm int}(t)/N_{\rm pix}$, where $N_{\rm pix}$ is the number of grid pixels associated with the corresponding emitting annulus of the accretion disk. A pixel is included in $N_{\rm pix}$ only if more than one-third of its area is covered by the annulus; this choice has a negligible effect on the results. The extinction map is defined as $E(B-V)(x_i,y_j,t) = E_{\rm peak} \exp\left[-\left((x_i-x_c(t))^2 + (y_j-y_c(t))^2\right)/\sigma_{\rm cloud}^2\right]$, where $E_{\rm peak}$ is the central color excess and $\sigma_{\rm cloud}$ characterizes the spatial extent of the cloud. We adopt $E_{\rm peak} = 0.5$ and $\sigma_{\rm cloud} = 2.0$. The transmission is computed using the \citet{1999PASP..111...63F} extinction law, with coefficients $k_\lambda$ derived from \texttt{FM\_UNRED}, such that $T_\lambda(x_i,y_j,t) = 10^{-0.4\, k_\lambda\, E(B-V)(x_i,y_j,t)}$. We note that alternative extinction curves often adopted for AGNs (e.g., SMC-like laws, \citet{Hopkins2004}) would mainly rescale the optical attenuation and are therefore unlikely to qualitatively affect the lag behavior. The observed flux is then obtained by summing over all emitting pixels, $F_{\rm obs}(t) = \sum_{i,j} F_{\rm pixel}(i,j,t)\, T_\lambda(x_i,y_j,t)$, thereby accounting for both geometric overlap and spatially varying extinction. Finally, the flux is converted back to magnitudes as $m_{\rm obs}(t) = -2.5 \log_{10}[F_{\rm obs}(t)/F_0]$.

\begin{figure*}[!ht]
		\centering
		\includegraphics[width=0.9\textwidth]{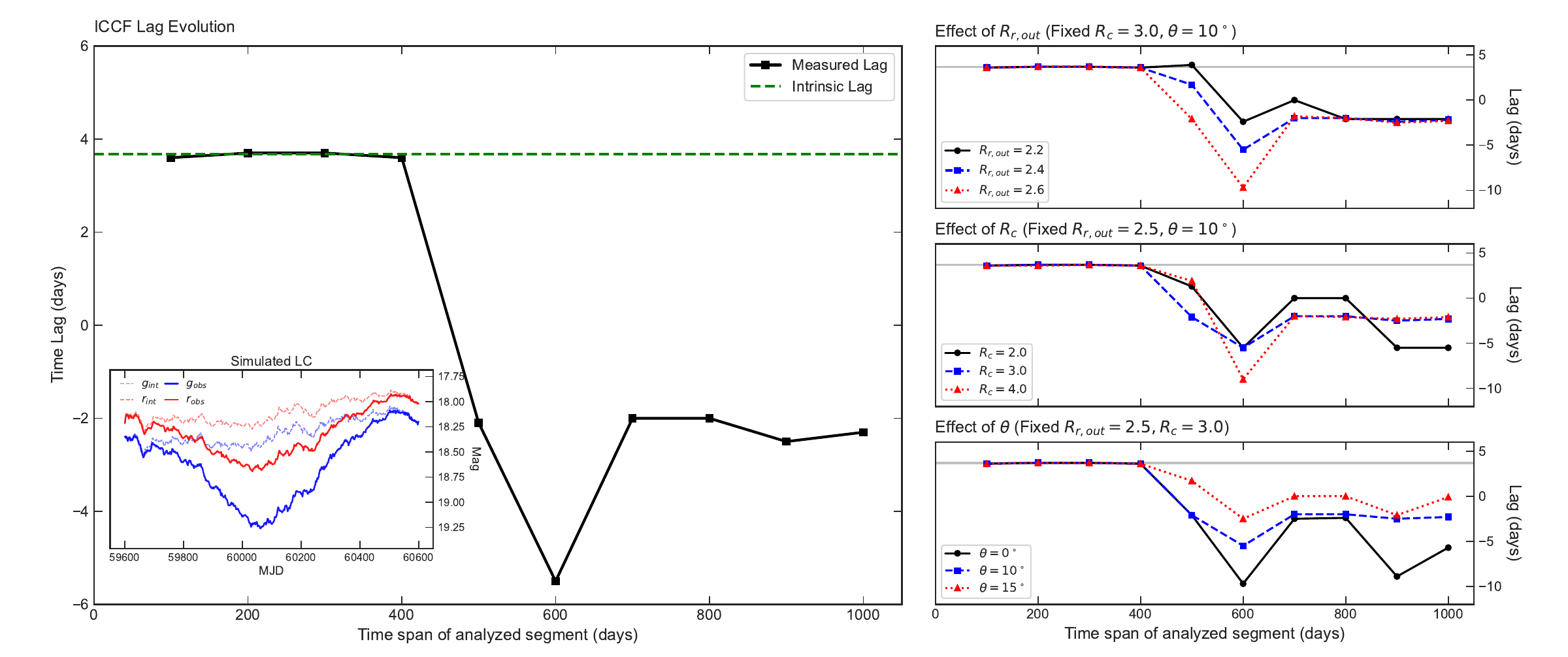} 
		\caption{Left panels: Evolution of the measured inter-band time lag as a function of the time span of the analyzed light-curve segment, starting from the onset of the simulated obscuration event. The green dashed line indicates the intrinsic positive lag ($+3.68$ days). The black points show the measured ICCF lag. The inset panel in the bottom-left corner displays the simulated intrinsic (dashed) and obscured (solid) light curves. Right panels: The effects of varying the $r$-band outer radius $R_{r,out}$ (top), the cloud radius $R_c$ (middle), and the crossing angle $\theta$ (bottom). The horizontal dashed gray line represents the input intrinsic time lag between the bands.}
		\label{fig:results}
	\end{figure*}

	The simulation results are shown in left panel of Figure~\ref{fig:results}. As the obscuration event progresses ($t \gtrsim 500$~days), the measured ICCF lag shifts from the intrinsic positive reverberation value to negative values, reaching a minimum around $t \sim 600$~days when the outer $r$-band region is most affected. After the cloud uncovers the central $g$-band region, the lag increases but remains negative because the extended $r$-band emission continues to respond earlier. The intrinsic lag is recovered only after the cloud exits the system. The parameter sensitivity, shown in Figure~\ref{fig:results} (right panels), supports an obscuration-driven origin for the lag inversion. Increasing the outer radius of the $r$-band ($R_{r,\rm out}$) and the cloud radius ($R_c$) enhances the negative lag by increasing the obscuration cross-section. Conversely, increasing the inclination angle ($\theta$) weakens the effect by reducing the effective covering of the disk.

	An additional factor influencing the measured lag is the degree of radial overlap between the $g$- and $r$-band emission regions. A larger overlap reduces the effective spatial separation between the two bands, thereby weakening the differential impact of obscuration. As a result, the induced negative lag is suppressed and the measured lag remains closer to the intrinsic positive value. Conversely, a smaller overlap enhances the radial stratification, allowing the cloud to intercept the outer $r$-band–dominated region earlier relative to the more compact $g$-band emission, which strengthens the negative lag signal. In addition, the numerical resolution of the Cartesian grid affects the precision of the measured lag. In our simulations, a grid size of $400 \times 400$ is sufficient to ensure convergence, and further increases in resolution do not produce significant changes in the measured lag. We note that the present calculation is based on a single realization for a specific set of stochastic variability parameters. A more comprehensive exploration of the parameter space, including different combinations of $\sigma$, $\tau$, $E_{peak}$, and $\sigma_{cloud}$, will be presented in future work.
	
	Our obscuration model is a simplified geometric description of the observed lag. Order-of-magnitude estimates of the cloud properties are provided in Appendix~\ref{apenc}, showing that the proposed scenario is physically plausible for dusty gas in AGNs. We do not attempt to draw firm conclusions about the physical origin, formation mechanism, or detailed spatial structure of the obscuring clouds. We suggest that the anomalous negative lag observed in SDSS J0837 arises from the transit of a dust cloud across the line of sight, which preferentially occults the outer $r$-band emitting regions. In this picture, the obscuration is expected to be transient; the cloud may move out of the line of sight on a timescale of a few years, potentially around 2027, allowing the light curve to recover toward its pre-obscuration baseline. Obscuration remains a viable explanation for a small subset of ``changing-look'' AGNs \citep{2023NatAs...7.1282R, 2025ApJ...988...25A}. Our results are therefore consistent with the possibility that line-of-sight obscuration can still contribute to transient variability in some systems. These results highlight that extrinsic geometric effects, rather than intrinsic changes in the accretion disk, may provide a viable explanation for the observed short-term lag anomalies and variability in AGNs.
	

	\section{Conclusion}\label{con}
	
	We report the detection of a short-timescale negative lag of approximately $-1.60$ days in the optical continuum of SDSS J0837 between the ZTF g- and r-bands during the declining phase of SDSS~J0837. This short time-scale negative inter-band lag, independently confirmed by both the \texttt{JAVELIN} and ICCF methods, together with the positive lag measured during the stabilizing phase, indicates a clear phase-dependent lag reversal. We therefore propose a moving dust-cloud obscuration model, in which the cloud transiting the line of sight preferentially attenuate the $r$-band emission and naturally produce the observed lag inversion.

	\begin{acknowledgements}
		Zhang gratefully thanks the kind financial support from GuangXi University, the HangJi Action Plan under the Guangxi Science and Technology Program 2026GXNSFDA00640018, the kind grant support from NSFC-12373014 and NSFC-12173020, the support from the Guangxi Talent Programme (Highland of Innovation Talents), and the support from the Bagui Scholars Programme (W X G., GXR-6BG2424001). This manuscript has made use of the data from the ZTF survey. The ZTF website is \url{http://irsa.ipac.caltech.edu/} . 
		
	\end{acknowledgements}
	
	\bibliographystyle{aa}
	\bibliography{aa}
	
	\clearpage 
	
	\appendix
	
	\section{Posterior distribution of the transfer-function width}
\begin{figure}
		\centering
		\includegraphics[width=\columnwidth]{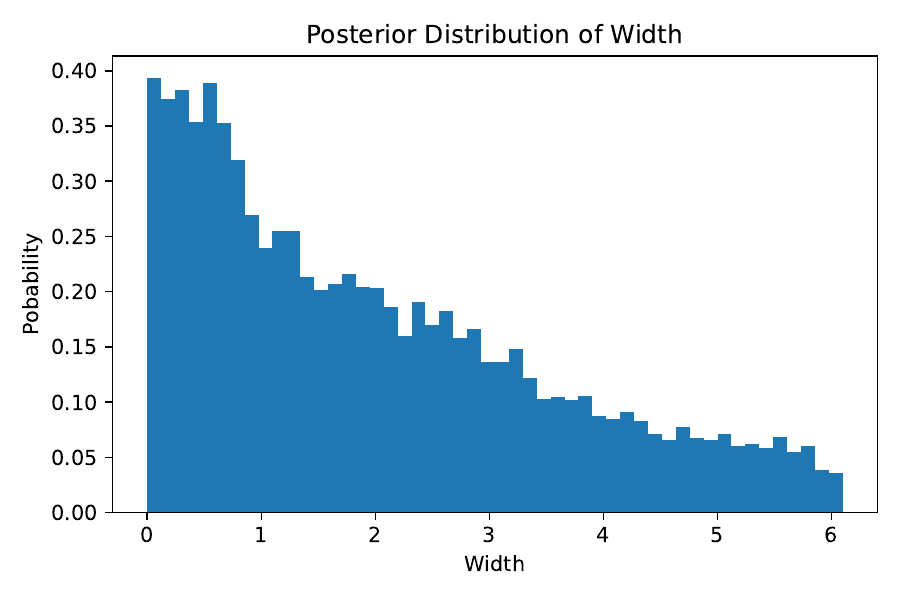} 
		\caption{Posterior probability distribution of the transfer-function width obtained from the \texttt{JAVELIN} RM analysis.}
		\label{fig:width}
	\end{figure}

	Figure~\ref{fig:width} shows the posterior probability distribution of the transfer-function width obtained from the JAVELIN analysis. The distribution is strongly peaked near zero, indicating that the response between the g- and r-band light curves is consistent with a narrow transfer function.
	
	\section{Sampling properties and their effects on lag measurements}
	\label{apenb}
	
	\subsection{Cadence properties of the ZTF light curves}
	We examine the cadence distributions of the g- and r-band light curves, defined as the time intervals between consecutive observations.
	\begin{figure*}
		\centering
		\includegraphics[width=\textwidth]{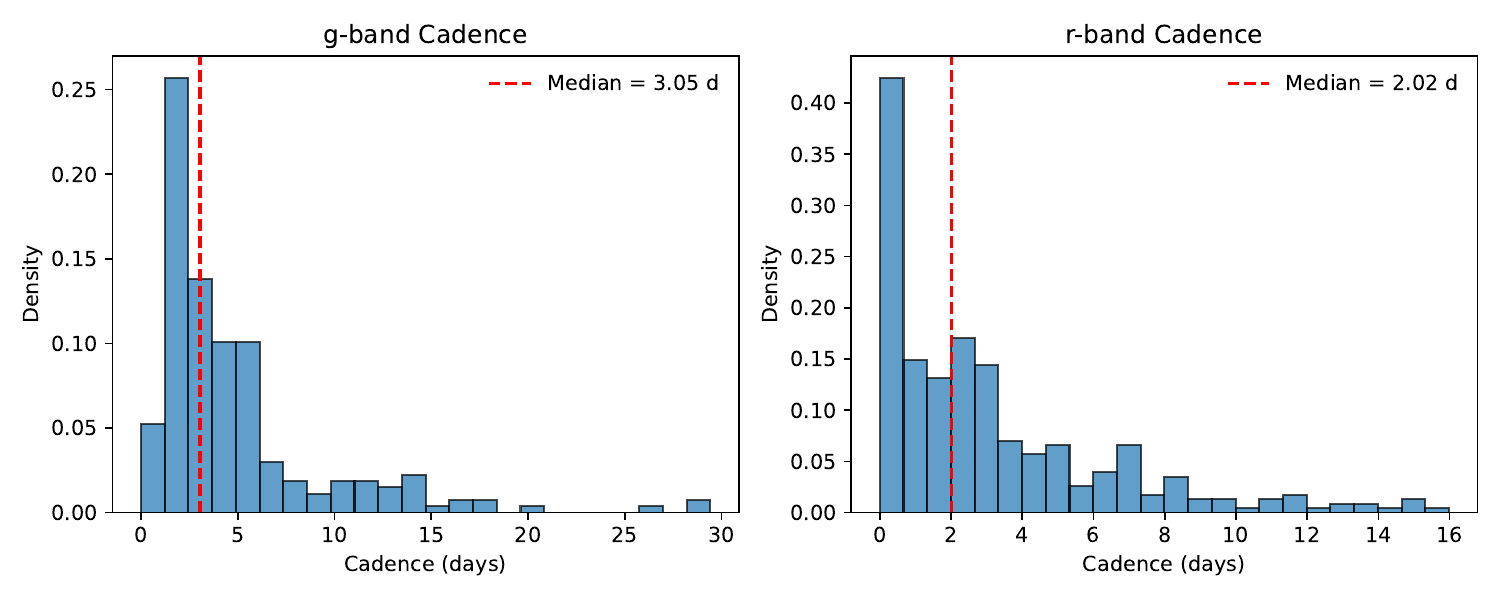} 
		\caption{Cadence distributions of the ZTF light curves. Left: histogram of the $g$-band cadence. Right: histogram of the $r$-band cadence. In each panel, the red dashed line indicates the median cadence (3.05 days for $g$-band, 2.02 days for $r$-band).}
		\label{cadence}
	\end{figure*}
	
	\begin{figure*}
		\centering
		\includegraphics[width=\textwidth]{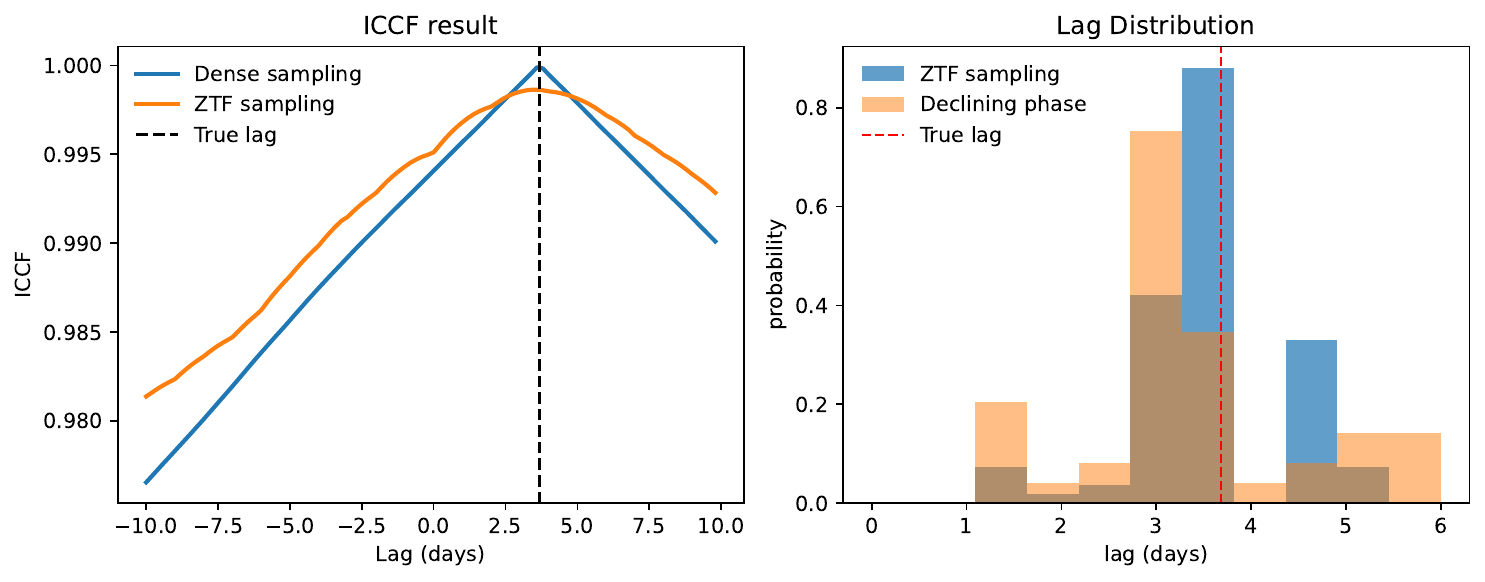}
		\caption{Left: ICCF results for a representative realization. The blue and orange curves show the results obtained with dense sampling and ZTF-like sampling, respectively. The black dashed vertical line marks the true lag. Right: Distributions of recovered lags from 200 simulations. The blue and orange histograms correspond to the ZTF-sampling case and the declining-phase case, respectively. The red dashed vertical line indicates the true lag.}
		\label{simu_re}
	\end{figure*}
	As shown in Fig.~\ref{cadence}, the median cadence is approximately 3.05 days in the g-band and 2.02 days in the r-band. Most observations are separated by a few days, while the distributions exhibit extended tails toward longer intervals, indicating the presence of occasional large gaps.
	
	This confirms that the light curves are characterized by moderately sparse and uneven sampling, which may affect the reliability of lag measurements and therefore motivates the simulation analysis presented below.
	
	\subsection{Simulation of sampling effects}
	To assess whether the observed negative lag could be induced by the sampling pattern, we perform simulations based on a CAR(1) process.
	
	We first generate a continuous mock g-band light curve with a time resolution of 0.1 days using CAR(1) process. The corresponding r-band light curve is constructed by introducing a time delay of 3.68 days, ensuring that the intrinsic lag is positive.
	
	The simulated light curves are then resampled using the actual ZTF observation times, thereby preserving both the cadence and the large observational gaps. Observational noise comparable to the ZTF photometric uncertainties is also included.
	
	We compute the lag using the ICCF method. As shown in Fig.~\ref{simu_re} (left panel), the lag recovered from densely sampled light curves is consistent with the input value. After applying the ZTF sampling, the ICCF becomes broader, but the peak remains located at positive lag.
	
	We repeat the simulation for 200 independent realizations. The distribution of recovered lags is shown in Fig.~\ref{simu_re} (right panel). All recovered lags remain positive, although the scatter increases under sparse sampling conditions. Similar behavior is found when only the declining phase is considered.
	
	
	These results suggest that, for the ZTF sampling considered here, sparse sampling, large gaps, and observational noise could be unlikely to produce spurious negative lags, and therefore may not fully account for the observed lag inversion in this dataset.
	
	\section{Physical properties of the obscuring cloud}
	\label{apenc}
	\subsection{Density estimate}
	We first estimate the characteristic density of the obscuring cloud based on its optical depth. 
	For a dusty medium, the optical depth can be expressed as
	\begin{equation}
		\tau \sim \kappa_{\mathrm{dust}} \rho 2R_c,
	\end{equation}
	where $\kappa_{\mathrm{dust}}$ is the dust opacity, $\rho$ is the mass density, and $R_c$ is the cloud radius.
	
	Assuming a marginal optical depth of $\tau \sim 1$ in the optical band, a typical dust opacity of $\kappa_{\mathrm{dust}} \sim 2.5 \times 10^{2}\ \mathrm{cm^2\,g^{-1}}$, and a cloud radius of $R_c \sim 6$ light-days, we obtain
	\begin{equation}
		\rho \sim \frac{\tau}{2\kappa_{\mathrm{dust}} R_c} \sim 10^{-19}\ \mathrm{g\,cm^{-3}}.
	\end{equation}
	
	This corresponds to a particle number density of $n \sim 10^{4}\ \mathrm{cm^{-3}}$, assuming a mean particle mass appropriate for a partially ionized gas. Although this density is lower than typical values for the BLR, it is consistent with dusty gas located in the outer BLR or participating in radiatively driven outflows, as suggested in the FRADO scenario \citep{2011A&A...525L...8C, Naddaf2021, 2022A&A...663A..77N}. This value also lies within the expected range for dusty gas in AGN environments.
	
	\subsection{Cloud mass}
	The total mass of the obscuring cloud can be estimated from its density and size. 
	Assuming a spherical geometry, the cloud mass is given by
	\begin{equation}
		M_c \sim \frac{4}{3}\pi R_c^3 \rho.
	\end{equation}
	
	Substituting $R_c \sim 6$ light-days and $\rho \sim 10^{-19}\ \mathrm{g\,cm^{-3}}$, we obtain
	\begin{equation}
		M_c \sim 10^{-3} - 10^{-2}\ M_{\odot}.
	\end{equation}
	
	This relatively small mass indicates that the obscuration is produced by a localized and transient structure, rather than a large-scale component of the torus.
	
	\subsection{Cloud velocity and physical location}
	The characteristic velocity of the cloud can be estimated from its crossing timescale. 
	Assuming a crossing time of $\sim 1000$ days over a spatial scale of $\sim 6$ light-days, we obtain
	\begin{equation}
		v_{\mathrm{cross}} \sim \frac{R_c}{t} \sim 10^{3}\ \mathrm{km\,s^{-1}}.
	\end{equation}
	
	If interpreted as Keplerian motion,
	\begin{equation}
		v_{\mathrm{K}} \sim \left(\frac{GM_{\mathrm{BH}}}{R}\right)^{1/2},
	\end{equation}
	the inferred velocity $v_{\mathrm{cross}} \sim 10^{3}\ \mathrm{km\,s^{-1}}$ would correspond to radii much larger than $\sim 100\,R_g$ for a black hole mass of $M_{\mathrm{BH}} \sim 10^{9}\ M_{\odot}$. 
	This suggests that the cloud motion is more plausibly associated with gas located in the outer BLR or near the dust sublimation region, rather than gas orbiting at $\sim 100\,R_g$.
	
	The inferred velocity lies within the low-to-intermediate range of BLR kinematics, although it is somewhat higher than typical values associated with the torus. 
	This is consistent with a scenario in which the cloud is located near the outer BLR or the dust sublimation radius, where the BLR is thought to form, as in the FRADO framework.
	
	For high-luminosity AGNs such as SDSS~J0837, the dusty structure is expected to be relatively compact in units of $R_g$, following the scaling relation $R_{\mathrm{torus}} \propto L_{\mathrm{bol}}^{1/2}$. 
	Given that $L_{\mathrm{bol}}$ approximately scales with $M_{\mathrm{BH}}$ within a narrow Eddington ratio range, this implies $R_{\mathrm{torus}}/R_g \propto M_{\mathrm{BH}}^{-1/2}$, leading to relatively higher characteristic orbital velocities. 
	As a result, transient obscuration by dusty clouds may be more readily observable in AGNs with large black hole masses.
	
\end{document}